\documentclass{kluwer}    
 
\newdisplay{guess}{Conjecture}
\usepackage{epsfig}
\def\deg{$^{\circ}\,$}

\begin{document}

\begin{article}
\begin{opening}
\title{Physical Morphology of Galaxies using Asymmetry}
\author{Christopher J. \surname{Conselice}}
\runningauthor{C.J. Conselice}
\runningtitle{Physical Morphology of Galaxies using Asymmetry}
\institute{University of Wisconsin-Madison}

\date{October 1, 1999}

\begin{abstract}
  We demonstrate in this paper the use of asymmetry in conjunction with the
integrated (B-V) color of galaxies for physical morphological purposes.  We
show how color-asymmetry diagrams can be used to distinguish between various
types of galaxies, including ellipticals, late and early type disks,
irregulars and interacting/merging galaxies.  We also show how asymmetry
can be used to help decipher the morphology of nearby starbursts, and galaxies
in the Hubble Deep Field.

\end{abstract}

\end{opening}

\section{Introduction}

      Observable properties of astrophysical objects increases our knowledge
of them; the more properties, the better we are able to make characteristics.
In astronomical imaging, or spectroscopy, the extended structure of galaxies
offers many advantages, most of which have not been thoroughly explored.
Extragalactic studies have an advantage over stars in this regard, since a 
stellar image contains only magnitude and color information.   For this 
reason, classification of stars has been based on their spectrum.  However, an 
image of a galaxy is full of morphological detail that can be effectively 
exploited.

   A host of these photo-morphologic properties can be determined,
that include, but are not limited to: magnitudes, colors, color
gradients, morphology, light concentrations, surface brightness, and 
asymmetry.  The last of these, asymmetry is one of the hardest of these 
parameters to measure, but is a powerful tool that can be used to determine 
both morphological and physical parameters of galaxies.

      Asymmetry can be measured in a several different ways.  These
include: Fourier component analysis (e.g. Rix \& Zaritsky 1995), searches for 
azimuthal deviations (e.g. Kornreich et al. 1998), or
by the use of radial 180\deg rotations (Conselice 1997; Conselice, Bershady,
\& Jangren 2000).   The asymmetry algorithm used in this paper is described 
in detail in Conselice et al. (2000).  Improvements of
this method over previous asymmetry computations (e.g. Abraham et al. 1996;
Conselice 1997) include a robust method of finding the center of rotation
which effectively iterates to find the minimum asymmetry, as well as using
a standard non-isophotal radius. 

  How asymmetry is useful as a morphological diagnostic is the focus
of this paper;  we will show how this parameter can be used in several
extragalactic environments.

\section{Nearby Galaxy Color-Asymmetry Diagram}

\begin{figure}
\centerline{\epsfig{file=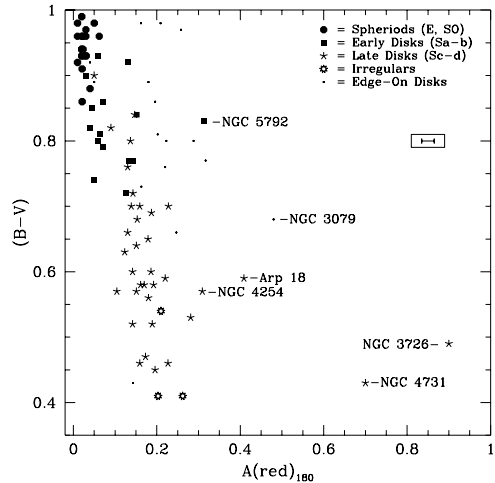,width=16pc}
\caption{Color-Asymmetry Diagram for Nearby Galaxies. Labeled galaxies
are interactions/mergers.}
}
\end{figure}

   Comparing asymmetry with the color of a galaxy can reveal important physical
and morphological information.   Shown in Figure 1 is the color-asymmetry
diagram from a sample of 113 galaxies taken from the Frei et al. (1996)
catalog.  These 113 galaxies sample all Hubble types, including inclined
and peculiar galaxies.

     There are several features in Figure 1 worth noting.  Most of the
galaxies fall along a fiducial sequence that expresses the correlation
between color and asymmetry.  Most galaxies get bluer that as they get more
asymmetric.  This is true for all Hubble-Types from ellipticals, through
spirals to irregular galaxies.

   An equally important property of this diagram are the galaxies that do not
fall on or near the fiducial sequence.  Some of these are galaxies that
are inclined, and therefore have a morphology driven by projection effects.
The galaxies in Figure 1 that are inclined are noted as tiny dots. 

   The galaxies in Figure 1 that are too asymmetric for their color, but 
are not inclined show evidence for recent interactions and mergers.
Galaxies considered from other considerations as an 
interaction/merger are label with their names.  Most of these
are significantly displaced from the fiducial color-asymmetry sequence. The 
power of the color-asymmetry diagram is this ability to automatically
sort out in a sample of galaxies, which ones are likely the result of
an interaction/merger. 

\begin{figure}
\centerline{
\epsfig{file=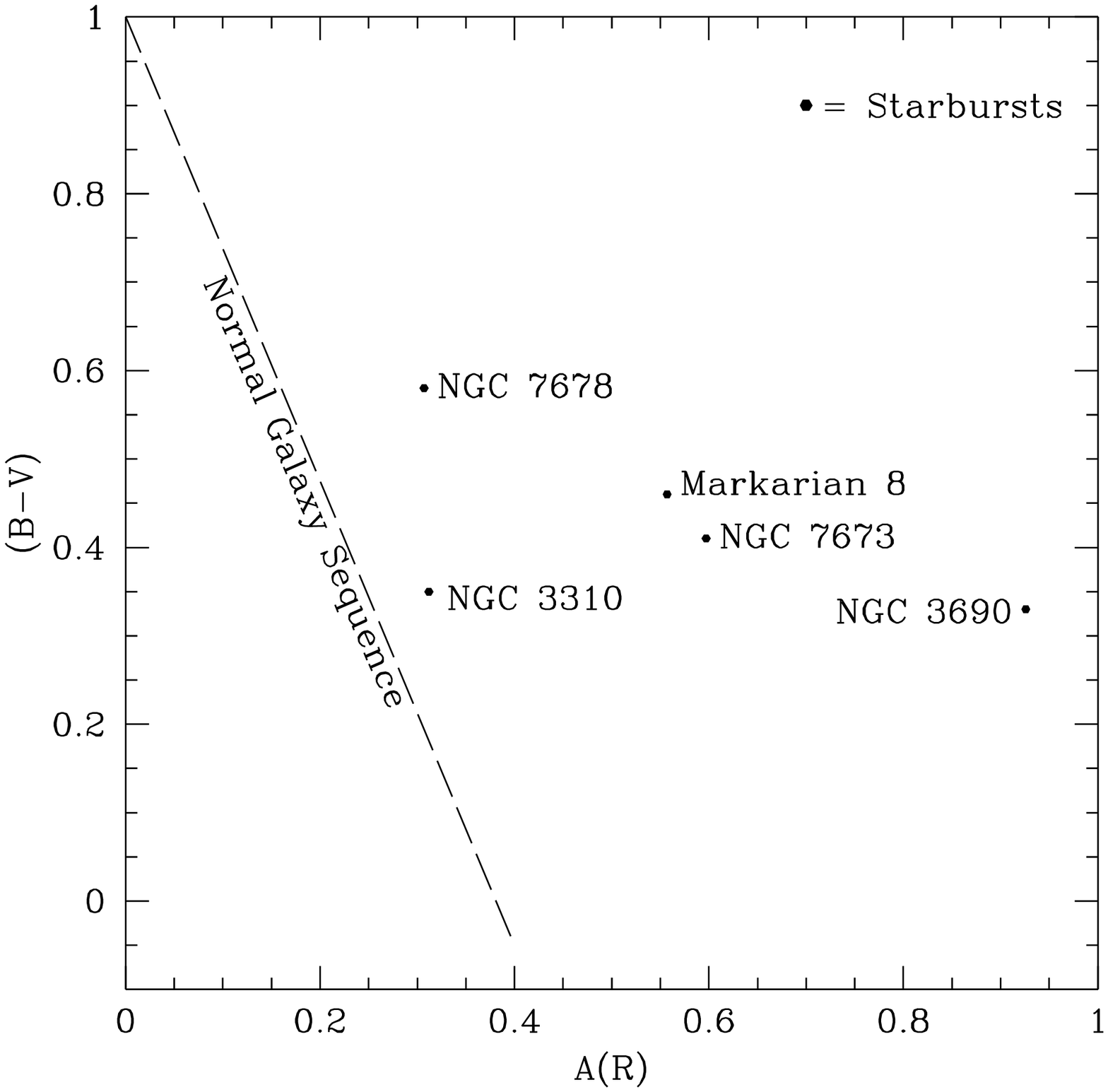,width=12pc}~A
\epsfig{file=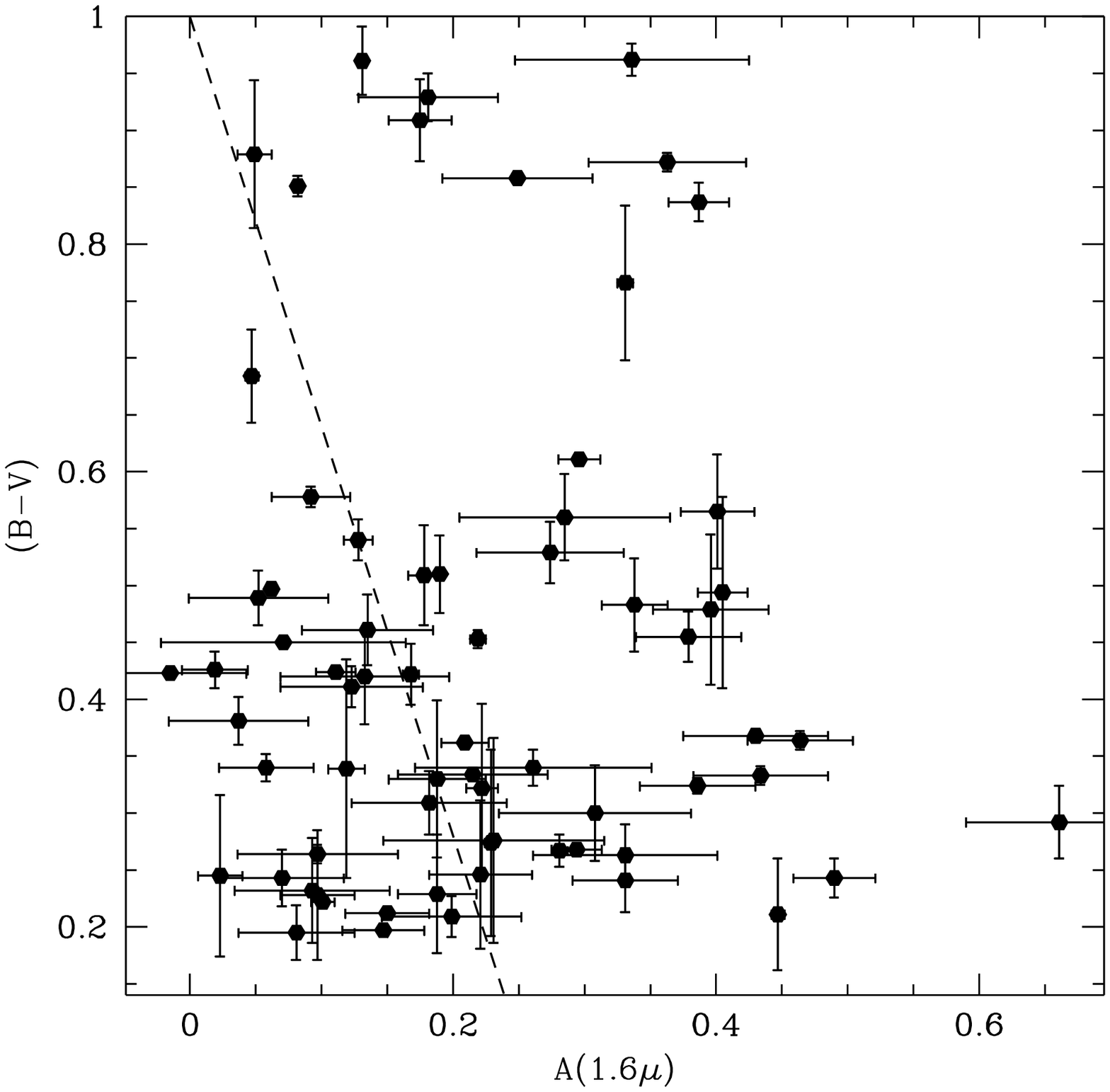,width=12pc}~B.
\caption{A) Color-Asymmetry Diagram for Starbursts  B) Color-Asymmetry Diagram
for galaxies in the HDF}
}
\end{figure}

\section{Physical Morphology at High Redshift}

   The power of asymmetry also lies in its ability to determine the history 
of  a galaxy, and whether or not it has undergone a recent interaction.  This 
is important for understanding galaxies in a host of environments, but can 
also be useful for determining possible trigger mechanisms for starbursts 
and other morphologically peculiar galaxies.  

  In Conselice et al. (2000; in prep) we show that by using
the color and asymmetry of starburst galaxies, it is possible to determine
if the triggering mechanism for starburst involved a galaxy interaction/merger,
or if the likely triggering mechanism is a bar-instability or another
internal event.  Figure 2a shows the color-asymmetry diagram for a sample of
nearby starbursts, most are consistent with a merger/interaction, except for 
NGC 3310 which was probably triggered by a bar-instability. We can 
quantitatively compare nearby starbursts with high redshift
galaxies to determine if these two populations are similar, as has been
suggested.

 To compare nearby and distant galaxies, we compute the asymmetries of 
galaxies in the NICMOS images of the Hubble Deep Field in the H band 
(1.6$\mu$).  We also use k-corrected (B-V) colors of these galaxies, based on 
6 color photometry based on  the technique of Bershady (1995).

   Previous observations of galaxies at high redshift indicate that a large 
fraction are undergoing star formation (Steidel et al. 1996).  These galaxies 
are similar in structure to nearby starburst galaxies (e.g. Hibbard \& 
Vacca 1997; Giavalisco et al. 1996) as well as more detailed morphological 
properties (Conselice et al. 2000; AJ in press).  However, are both 
populations undergoing mergers/interactions, and what has triggered the 
star formation?

   Star formation can occur by various methods, including interactions 
with other galaxies, as well as internal processes such as bar 
instabilities.  The likely mechanism behind the triggering of a starburst 
event can be determined by finding its position on a color-asymmetry diagram.
We must be cautious however, when attributing asymmetries in high redshift
galaxies for an interaction, or merger.  The peculiar forms seen in these
galaxies could in some cases result from the initial creation of these 
galaxies.  However, Figure 2b does show that the majority of galaxies at
high redshift have colors and asymmetries that are not consistent with
normal nearby galaxies, even irregulars.  The galaxies seen at high redshift
are therefore either going through an interaction/merger or are in the
process of stabilizing from their initial creation. 

\begin{acknowledgements}

  I thank Matt Bershady, Mark Dickinson and Jay Gallagher, and the NICMOS
HDF team, for their many
contributions and ideas to this work. The support of 
Professor D. Block, and the sponsors of this conference, the Anglo American 
Chairman's Fund, and SASOL is appreciated.
This work was supported by the NASA/Wisconsin Space Grant 
Consortium, and by a Grant-in-Aid of Research from the National Academy
of Sciences, through Sigma Xi.

\end{acknowledgements}
\theendnotes

\end{article}
\end{document}